\documentclass[final,1p,times]{elsarticle}

\usepackage{amssymb}
\usepackage{amsmath}

\usepackage{multirow}
\usepackage{rotating}  
\usepackage{tabularx}

\journal{Applied Energy}

\begin{document}

\begin{frontmatter}



\title{A Deep Learning-Based Method for Power System Resilience Evaluation}


\author{Xuesong Wang} 
\ead{xswang@wayne.edu}
\author{Caisheng Wang\corref{cor}}
\ead{cwang@wayne.edu}
\cortext[cor]{Corresponding author: Caisheng Wang.}

\affiliation{organization={Department of Electrical and Computer Engineering, Wayne State University},
            addressline={42 W Warren Ave},
            city={Detroit},
            postcode={48202},
            state={MI},
            country={USA}}
\fntext[ack]{This material is based upon work supported by the Department of Energy, Solar Energy Technologies Office (SETO) Renewables Advancing Community Energy Resilience (RACER) program under Award Number DE-EE0010413. Any opinions, findings, conclusions, or recommendations expressed in this material are those of the authors and do not necessarily reflect the views of the Department of Energy.}

\begin{abstract}
Power system resilience is vital to modern society, as outages caused by extreme weather can severely disrupt communities. Existing statistical and simulation-based methods for resilience quantification are either retrospective or rely on simplified physical models, limiting their applicability. This paper proposes a deep learning-based framework that integrates historical outage and weather data to predict event-level resilience, measured using the resilience trapezoid method. The trained model is then applied to a benchmark weather dataset to estimate regional resilience, with optional socioeconomic and demographic factors incorporated as weighting terms when policymakers wish to emphasize the needs of specific population groups. The effectiveness of the framework is first validated on simulated outage records, showing strong agreement between predicted and simulated resilience values. It is then applied to real historical outage data to assess the resilience of actual power systems. Beyond evaluation, the results can guide targeted investments in distributed energy resources to improve resilience in vulnerable regions.
\end{abstract}

\begin{keyword}
deep learning \sep evaluation \sep power system \sep resilience.


\end{keyword}

\end{frontmatter}



\section{Introduction}

\begin{figure}
    \centering
    \includegraphics[width=1\linewidth]{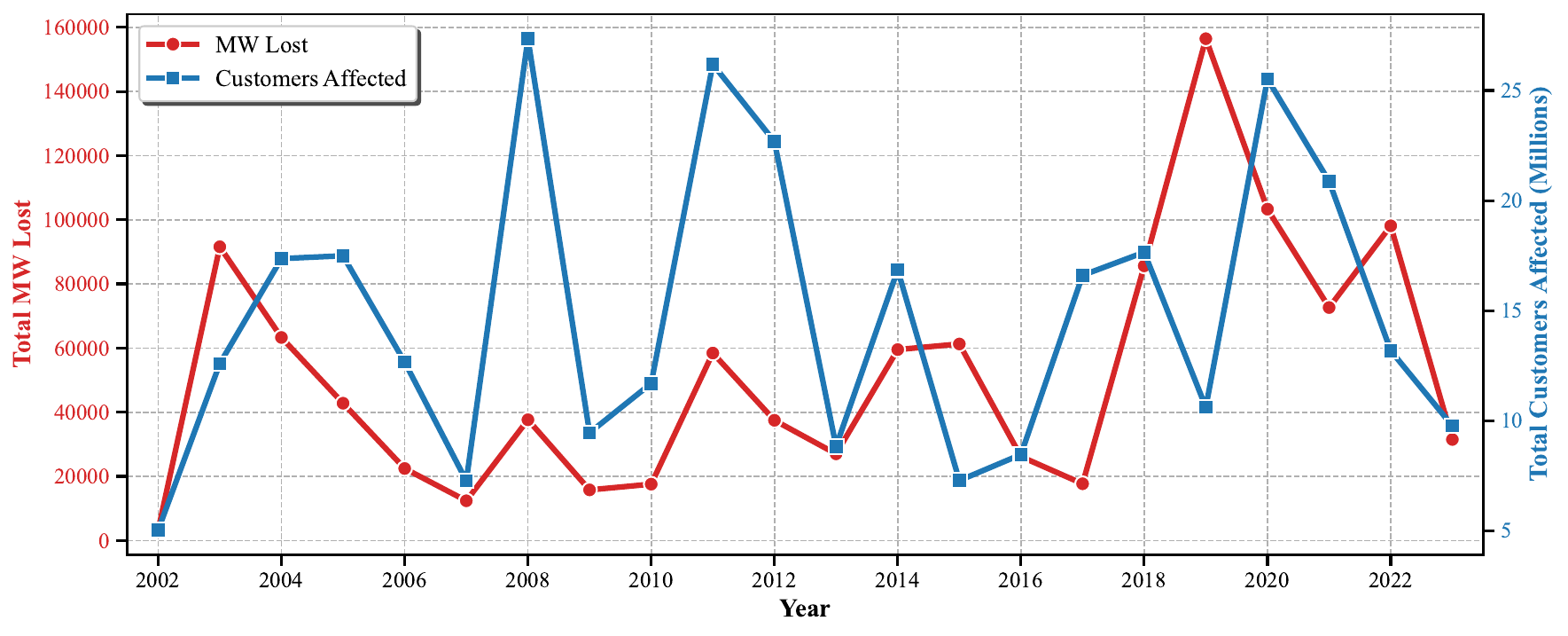}
    \caption{Trends in annual MW lost and customers affected from DOE OE-417 summaries (2002-2023).}
    \label{fig:doe417_outage_trends}
\end{figure}

Reliable access to electricity is fundamental to modern society, and large-scale power outages can cause severe economic losses, social disruption, and even life-threatening consequences. In the United States, approximately 80\% of major outages reported between 2000 and 2023 were triggered by weather-related events \cite{R53}. The frequency of such events has also increased, with twice as many weather-induced outages occurring during 2014-2023 compared to 2000-2009 \cite{R53}. The U.S. experienced 27 billion-dollar weather and climate disasters in 2024—the second-highest annual count on record—many of which caused widespread, multi-day power interruptions \cite{R73}. Trends in annual MW lost and customers affected from DOE OE-417 summaries (2002-2023) are shown in Fig.~\ref{fig:doe417_outage_trends}. This growing trend highlights the urgent need to better understand how power systems perform under disruptive conditions and to develop systematic methods for evaluating their ability to withstand and recover from extreme events.

Power system performance under disruptive conditions is commonly discussed in terms of two related but distinct concepts: reliability and resilience \cite{R11}. Reliability emphasizes both adequacy—the ability to continuously supply electricity to end-users—and security, the capacity to withstand routine disturbances such as component failures or minor weather events \cite{R11,R54,R55,R56}. These considerations primarily address high-probability, low-impact (HPLI) events. In contrast, resilience focuses on the system's ability to absorb, adapt to, and recover from low-probability, high-impact (LPHI) disruptions, including hurricanes, floods, earthquakes, and cyber-physical attacks \cite{R11,R57}. While reliability captures the performance of power systems under everyday uncertainties, resilience highlights their robustness in the face of rare but catastrophic events.

Two main approaches have been developed to assess power system resilience: statistics-based methods and simulation-based methods \cite{R25}. Statistics-based approaches rely on historical outage records to construct indicators that summarize past system performance \cite{R7}. Because they do not require detailed information such as system topology or fragility models, they provide a straightforward retrospective view of how a system responded to disruptive events. However, the scarcity of LPHI events limits their scope: Most analyses focus on case studies of specific storms or hurricanes, and resilience comparisons across regions are difficult when similar disruptive events have not been observed.

Simulation-based approaches, by contrast, employ physical system models to evaluate performance under hypothetical scenarios derived from hazard models and fragility curves \cite{R1,R72}. Resilience in this setting is often quantified using the resilience trapezoid \cite{R26,R72} over multiple rounds of Monte Carlo simulation. These methods provide a controlled environment for fair comparisons across systems without requiring extensive outage records. Yet, their applicability is limited by the difficulty of obtaining accurate topology and component fragility data, and the simplifications used in hazard modeling may fail to capture the complexity of real-world events.

While physical system performance is a central concern, it does not fully capture the impacts of prolonged outages. Communities experience disruptions differently depending on their social and demographic characteristics. For example, a three-dimensional metric of social vulnerability—encompassing health, preparedness, and evacuation intention—has been proposed to evaluate how long-duration outages put different populations at risk \cite{R16,R46}. Similarly, an empirical study of outages following Hurricane Hermine in Tallahassee, FL, highlighted disparities across socio-demographic groups in their ability to cope with service interruptions \cite{R31}. More recently, \cite{R5} introduced a hybrid framework that integrates infrastructure and community factors into an Area Resilience metric (ARez), spanning energy, public health, natural ecosystems, socio-economic factors, and transportation. These studies underscore the importance of considering both physical and social dimensions when evaluating resilience, particularly when identifying vulnerable populations and planning targeted improvements.

In this work, we propose a deep learning-based framework for power system resilience evaluation that bridges the gap between statistics-based and simulation-based approaches. Leveraging historical outage and weather data, a deep neural network is trained to predict system performance as quantified by the resilience trapezoid \cite{R1}. To enable fair and consistent comparisons across regions, we construct a benchmark weather dataset against which resilience can be evaluated. The framework also allows, when desired, the optional incorporation of socio-economic and demographic weighting to reflect the vulnerabilities of specific population groups. In this way, the proposed method provides a flexible and scalable tool for resilience assessment that is data-driven, generalizable across systems, and adaptable to policy needs.

The remainder of this paper is organized as follows. Section \ref{section_2} reviews existing statistics-based and simulation-based approaches, as well as recent applications of deep learning to resilience evaluation. Section \ref{section_3} describes the proposed deep learning-based framework. Section \ref{section_4} presents two case studies to demonstrate its effectiveness. Section \ref{section_5} discusses limitations and directions for future work, and Section \ref{section_6} concludes the paper.

\section{Related Work}
\label{section_2}

\subsection{Statistics-based Methods for Resilience Evaluation}

Statistics-based methods typically assess power system resilience by constructing metrics from historical outage records or socio-economic indicators. One category of work develops indicator-based frameworks. For example, the authors of \cite{R2} proposed a framework for rural power systems that integrates 42 indicators across three dimensions: technical (21), social (8), and economic (13). While comprehensive, this framework was not demonstrated through a case study.

Another category leverages large-scale outage datasets. The Environment for Analysis of Geo-Located Energy Information (EAGLE-I) dataset, which records detailed outages across U.S. counties from 2014 to 2022 \cite{R52}, has supported a variety of resilience studies at county level. Using this dataset, \cite{R6} and \cite{R37} introduced metrics such as event duration, recovery duration, post-event duration, impact rate, recovery rate, and recovery-to-impact ratio, each modeled with a probability distribution. Although these studies did not consider weather, \cite{R35} combined EAGLE-I with National Weather Service records to evaluate resilience using metrics including Time Over Threshold (TOT), Area Under the Curve (AUC), and Time After the End of the event (TAE), enabling state- and county-level assessments across different types of weather events.

A third line of research utilizes the U.S. Department of Energy's Electric Emergency Incident and Disturbance Report (Form DOE 417). Using data from 2002 to 2016, the authors of \cite{R47} applied statistical trend tests to evaluate resilience metrics such as disruption rate, performance loss, and recovery time, and introduced a composite resilience measure. Building on this, \cite{R49} analyzed DOE 417 data from 2007 to 2018 to examine resilience factors across three dimensions: extrinsic disruptions, intrinsic system capacities, and the effectiveness of recovery.

Overall, statistics-based methods offer the advantage of grounding resilience analysis in real outage records, which ensures that the results reflect observed system behavior. However, their dependence on historical events limits generalizability: Different regions experience different hazards, making cross-regional comparisons challenging, particularly for rare low-probability, high-impact events.

\subsection{Simulation-Based Methods for Resilience Evaluation}

Simulation-based methods evaluate resilience by modeling the behavior of system components under hazardous conditions. A typical framework includes three elements: a system model, a hazard model, and a fragility model \cite{R27,R28,R38,R41,R44, R72}. For each simulated hazard, the performance of the system is tracked and resilience metrics are calculated \cite{R14}.

Fragility models can be developed either empirically or through physical modeling. For example, the fragility model in \cite{R7} was derived from outage records of Hurricane Hermine, and Monte Carlo simulations were used to evaluate two investment strategies: upgrading components and reducing restoration time. Alternatively, \cite{R38} investigated the fragility of distribution components under ice storms using physical modeling, accounting for four failure modes and the influence of wind attack angle.

Simulation approaches have also been extended beyond single networks. Some studies jointly analyzed reliability and resilience \cite{R18,R24}, while others examined interdependent infrastructures. For instance, resilience has been studied in combined electricity, gas, and heat networks \cite{R21,R29,R30,R42}, and in coupled power, gas, and water systems \cite{R45}. The InfraRisk platform proposed in \cite{R22} further integrates power, water, and transportation infrastructures for comprehensive risk assessment.

The main strength of simulation-based methods lies in their flexibility: By varying system, hazard, and fragility parameters, researchers can conduct controlled experiments and compare different designs or investment strategies. However, these models inevitably simplify complex real-world conditions. As a result, their predictions may diverge significantly from observed outcomes, limiting their value for practical decision-making when detailed and accurate system data are unavailable.

\subsection{Deep Learning Applications to Power System Resilience Evaluation}

Deep learning techniques have been increasingly applied to outage prediction and resilience studies. In \cite{R10}, a comprehensive review of machine learning applications to power system resilience was provided. Several works have focused on predicting outages directly. For instance, \cite{R15} proposed outage prediction models that identified relevant weather events and quantified their impact using data from the ERA5 and ERA5-Land reanalysis. Tree-based ensemble methods were explored in \cite{R20}, while \cite{R43} developed a hybrid framework that combined a feed-forward neural network with a Gated Recurrent Unit (GRU)~\cite{R62} to capture static and dynamic variables, respectively, and applied a multi-head attention mechanism to map disaster features to line interruptions. Using the EAGLE-I dataset, \cite{R51} further demonstrated the potential of machine learning models for outage forecasting. \cite{R69} incorporated socio-economic and infrastructure data in the model input and compared conditional models and unconditional models for power outage prediction.

Beyond prediction, deep learning has also been integrated into simulation-based resilience studies. In \cite{R1}, a Recurrent Neural Network was employed to generate synthetic typhoon scenarios for an IEEE 13-bus test system projected onto a southeastern coastal region of China, with resilience measured by load shedding under simulated disruptions. Similarly, \cite{R13} used a Graph Neural Network to identify critical nodes and links in distribution networks, supporting resilience analysis through structural vulnerability assessment.

While these studies highlight the growing role of deep learning, most existing efforts remain focused on outage prediction or simulation support rather than direct, data-driven quantification of resilience. In contrast, the approach proposed in this paper applies deep learning to predict resilience performance, as measured by the resilience trapezoid, under severe weather scenarios. This enables consistent evaluation across regions and bridges the gap between purely statistical and simulation-based methods.

\section{Methodology}
\label{section_3}

\subsection{Method Overview}
\label{section_3_1}

The proposed framework for evaluating power system resilience combines the strengths of both statistics-based and simulation-based approaches while avoiding their key limitations. Unlike simulation-based methods, it does not require detailed system topology or component fragility models. Instead, system behavior under severe weather scenarios is learned directly from historical outage and weather data using a deep learning model. Once trained, the model takes hazardous weather events and system descriptors as inputs to predict the resulting performance trajectory, which is then quantified through the resilience trapezoid method. Because the same benchmark set of hazardous events can be applied to multiple systems, the framework enables consistent and fair resilience comparisons across regions or utilities.

\begin{figure}[htbp]
    \centering
    \includegraphics[width=0.8\linewidth]{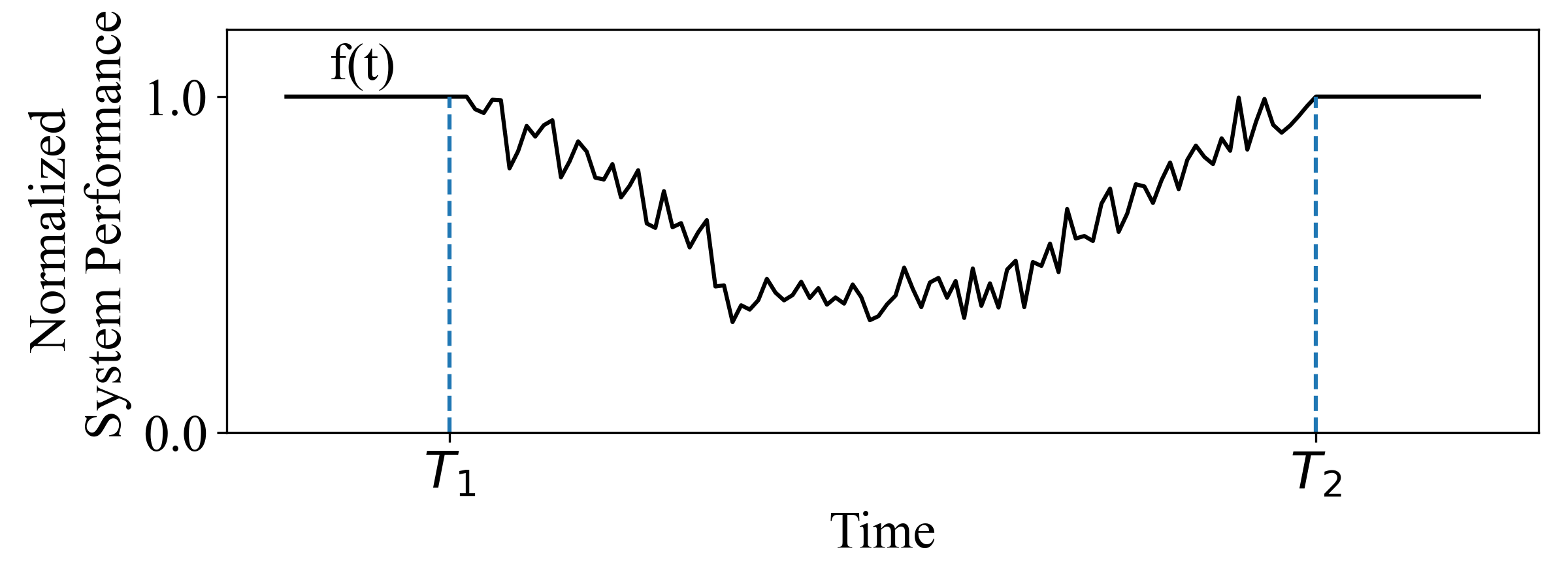}
    \caption{Demonstration of the system performance curve under a hazardous weather event. $f(t)$ is the normalized system performance curve. $T_1$ and $T_2$ are the start and end times of the outage event, respectively.}
    \label{fig:system_perf}
\end{figure}

In this framework, system performance is represented by the fraction of customers served, normalized by the total number of customers in the system; hence, the maximum performance is 1.0. When a severe weather event occurs, performance typically drops rapidly, remains at a reduced level while damaged components await repair, and then gradually recovers to full capacity, as illustrated in Fig.~\ref{fig:system_perf}. The area under this performance curve captures the system's ability to sustain service and recover, which is quantified using the resilience trapezoid method \cite{R26}. The resilience of system $k$ during outage event $i$ is defined in (\ref{eq:single_event_resilience}), where $f(t)$ denotes the normalized performance curve, and $T_1$ and $T_2$ represent the start and end times of the outage event, respectively. Note that $T_2$ differs from the end time of the weather event: The outage ends when the system is fully restored, which generally occurs later than the cessation of the hazardous weather.

\begin{equation}
    \label{eq:single_event_resilience}
    Rs_{i,k} = \frac{\int_{T_1}^{T_2}f(t)dt}{\int_{T_1}^{T_2}\textbf{1}dt} = \frac{\int_{T_1}^{T_2}f(t)dt}{T_2-T_1}
\end{equation}

To enable consistent and comparable resilience assessment across regions, a benchmark weather dataset is constructed containing a representative set of hazardous weather events. Each system under study is evaluated against this common benchmark, and its resilience is computed using (\ref{eq:unweighted_resilience}), where $Ru_k$ denotes the unweighted resilience of power system $k$, and $N_a$ is the number of hazardous events included in the benchmark dataset. The value of $Ru_k$ ranges from 0 to 1, with higher values indicating a more resilient system.

\begin{equation}
    \label{eq:unweighted_resilience}
    Ru_k=\frac{1}{N_a}\sum_{i=1}^{N_a}Rs_{i,k}
\end{equation}

While the unweighted resilience metric captures only the physical performance of the power system, different population groups may experience the consequences of power outages unequally. For example, individuals with disabilities or limited mobility are often more affected by service interruptions than others. To account for such disparities when desired, a weighted resilience metric is introduced, in which the unweighted resilience is adjusted by an exponential term incorporating social vulnerability factors across three dimensions—evacuation, preparedness, and health—as proposed in \cite{R16}. The formulation is shown in (\ref{eq:weighted_resilience}), where $Rw_k$ denotes the weighted resilience of power system $k$, $N_w$ is the number of socio-economic and demographic factors ($N_w = 15$ in this study), $W_{j,k}$ is factor $j$ for system $k$, and $\lambda$ is a coefficient controlling the influence of these factors ($\lambda = 1/3$ in this work). The 15 factors are listed in Table~\ref{table:weight_factors}. The socio-economic and demographic data were obtained from the U.S. Census Bureau's American Community Survey (ACS),
the Decennial Census, and the U.S. Department of Health and Human Services (HHS) emPOWER program. Each factor is normalized to ensure comparability; for example, household-related variables are divided by the total number of households in the region, while individual-related variables are normalized by population size.

\begin{equation}
    \label{eq:weighted_resilience}
    Rw_k=Ru_k^{[1+\lambda\sum_{j=1}^{N_w}W_{j,k}]}
\end{equation}

Because the unweighted resilience values range from 0 to 1 and all socio-economic and demographic factors are positive, larger vulnerability factors yield smaller weighted resilience values, reflecting the reduced adaptive capacity of more vulnerable regions. This formulation aligns with the principle that areas with higher social vulnerability tend to experience greater difficulty in coping with disruptions. The weighting mechanism also provides flexibility for policy-oriented analysis: Decision-makers can adjust or prioritize specific factors to emphasize the welfare of particular population groups, as demonstrated in Case Study~B.

\begin{table}[t]
\small
\centering
\caption{Socio-economic and demographic factors used for optional weighting of power-system resilience.}
\label{table:weight_factors}
\begin{tabularx}{\textwidth}{c|X|l}
\hline
Dimension & Factor & Source \\\hline
\multirow{3}{*}{Evacuation}
  & Households without vehicles & ACS (B08201) \\
  & Essential workers required on-site during outages & ACS (DP03) \\
  & People with disabilities & ACS (S1810) \\\hline
\multirow{7}{*}{Preparedness}
  & Younger adults & ACS (S0101) \\
  & Limited English proficiency & ACS (S1602) \\
  & Households with children & ACS (S1101) \\
  & Lower educational attainment & ACS (S1501) \\
  & Low-income households & ACS (DP03) \\
  & Elderly living alone & ACS (B25011) \\
  & Multifamily housing residents & ACS (S1101) \\\hline
\multirow{5}{*}{Health}
  & Older adults (65+) & ACS (S1810) \\
  & Young children (<5) & ACS (S1810) \\
  & Electricity-dependent medical equipment users & HHS emPOWER \cite{R63} \\
  & Nursing home residents & Decennial Census (P5) \\
  & Mobility limitations & ACS (S1810) \\\hline
\end{tabularx}
\end{table}

\subsection{Model Design}
\label{model_intro_section}

\begin{figure}[h]
    \centering
    \includegraphics[width=1\linewidth]{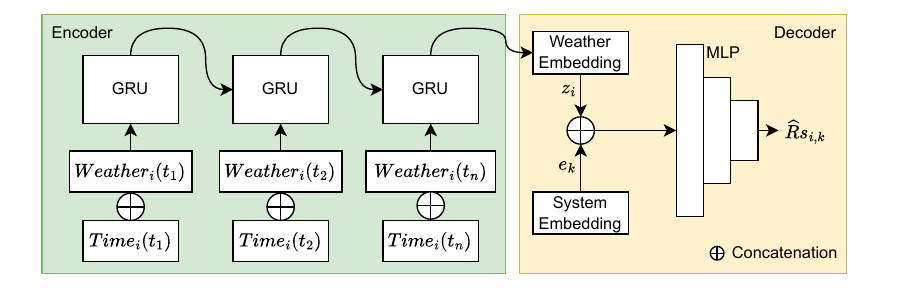}
    \caption{Model architecture for predicting $Rs_{i,k}$, the resilience of power system~$k$ under weather event~$i$, where $\hat{Rs}_{i,k}$ is the predicted value.}
    \label{fig:model_arch}
\end{figure}

A deep learning model is developed to estimate $Rs_{i,k}$, the resilience of power system~$k$ under weather event~$i$, as defined in~(\ref{eq:single_event_resilience}). The model serves as the computational core of the proposed framework and follows an encoder-decoder structure, as illustrated in Fig.~\ref{fig:model_arch}. The encoder extracts a compact representation of the multivariate weather scenario, while the decoder maps this representation—together with system-specific information—to the predicted resilience value $\hat{Rs}_{i,k}$.

\textbf{Encoder.}
The encoder processes sequential weather features associated with each hazardous event, such as wind speed, gust intensity, precipitation rate, and temperature. Because the temporal length of weather sequences varies among events, a Recurrent Neural Network (RNN) is employed to capture temporal dynamics. A Gated Recurrent Unit (GRU)~\cite{R62} is adopted for its computational efficiency and ability to handle long-term dependencies without vanishing gradients. Each input sequence is concatenated with an optional time-of-year embedding that encodes seasonal patterns, improving the model's generalization across different months. The encoder outputs a fixed-length weather embedding $z_i \in \mathbb{R}^d$ summarizing the hazard characteristics of event~$i$.

\textbf{Decoder.}
The decoder combines the weather embedding $z_i$ with a system embedding $e_k$ that characterizes the target power system. In this study, $e_k$ is implemented as a one-hot vector identifying each system, but it can be extended to incorporate additional descriptors such as customer density, energy-mix ratio, or infrastructure proxies. The concatenated vector $[z_i, e_k]$ is passed through a multi-layer perceptron (MLP) with nonlinear activation functions to produce the predicted resilience value $\hat{Rs}_{i,k}$:
\[
\hat{Rs}_{i,k} = f_{\theta}(z_i, e_k),
\]
where $f_{\theta}$ denotes the parameterized decoder network. The MLP depth, hidden dimension, and dropout rates are treated as tunable hyperparameters.

\textbf{Training and Optimization.}
Model parameters $\theta$ are optimized by minimizing the mean absolute error (MAE) between the predicted and observed event-level resilience values:
\[
\mathcal{L} = \frac{1}{N}\sum_{i,k}\big|\hat{Rs}_{i,k} - Rs_{i,k}\big|,
\]
where $N$ is the number of training samples. The network is trained using the Adam~\cite{R64} optimizer with learning-rate scheduling. A five-fold cross validation is utilized. The model with the lowest validation loss during training is used for testing. Dropout layers are applied within both GRU and MLP modules for regularization, and all input variables are normalized to the range $[0,1]$. Hyperparameter tuning—covering hidden sizes, layer numbers, learning rate, dropout, and weight decay—is performed via Ray Tune~\cite{R66} and Optuna~\cite{R67}, as detailed in Section~\ref{section_4}.

\textbf{Generalization and Adaptation.}
This encoder-decoder formulation enables the model to learn a generalized mapping between hazardous weather conditions and system performance. Once trained, it can predict resilience for unseen events or systems, provided their corresponding weather and system descriptors are available. In subsequent case studies (Section~\ref{section_4}), the same model architecture is instantiated under two distinct data regimes:
(1) a controlled synthetic dataset generated by a physics-based simulation framework for model validation, and
(2) real-world outage and weather records for statewide resilience evaluation.
Minor adjustments—such as omitting the time embedding when all hazards occur in summer—reflect data-specific configurations rather than architectural changes.

Compared with traditional outage-prediction models that focus on component failures or restoration times, the proposed model directly learns to predict the integrated resilience metric $Rs_{i,k}$, aligning its training objective with the overall system-level performance measure and bridging the gap between statistical and simulation-based resilience assessment approaches.

\section{Experiment Results}
\label{section_4}

Two case studies were conducted to validate and demonstrate the proposed resilience evaluation framework. In the first case study, a controlled validation was performed using synthetic data generated by a previously developed graph-based simulation framework~\cite{R72}. The simulation produced system performance records under thunderstorm-wind scenarios at the substation service area level, enabling a quantitative assessment of the model's predictive accuracy under well-defined conditions. In the second case study, the proposed method was applied to real-world power outage data at the county level in Michigan, USA to evaluate regional resilience and explore its implications for planning and policy. The datasets, model-training procedures, and resilience evaluation results for each case are described in the following subsections.

\subsection{Case Study A: Validation Using Synthetic Data}
\label{section_4_1}

\begin{figure}[thb]
    \centering
    \includegraphics[width=1.0\linewidth]{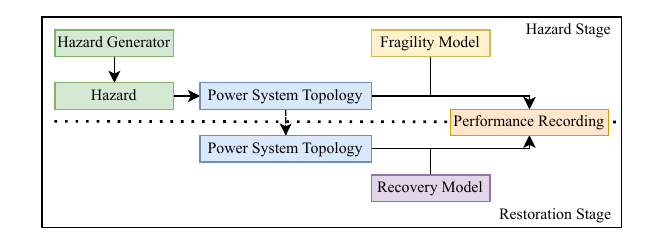}
    \caption{Architecture of the simulation framework in case study A.}
    \label{fig:simulation_framework}
\end{figure}

In this case study, the proposed deep learning framework was validated using synthetic data generated by a discrete-time simulation environment based on the graph-based resilience framework previously developed in~\cite{R72}. The simulation produces system performance records at the substation service area level under thunderstorm-wind scenarios. These physically simulated data provide a controlled benchmark for assessing whether the deep learning model can accurately learn the relationship between hazardous weather conditions and system resilience, measured using the resilience trapezoid. The simulation framework consists of four modules: a power system topology synthesizer, a hazard generator, a fragility model, and a recovery model. The process includes two sequential stages: the hazard stage and the restoration stage, as illustrated in Fig.~\ref{fig:simulation_framework}.

\subsubsection{Simulation Framework Overview}

The simulation operates in discrete hourly time steps and follows a Monte Carlo process to estimate resilience across multiple stochastic episodes. For demonstration purposes, only the fragility of power lines was considered in this study. Each episode is divided into two stages: the hazard stage and the restoration stage. During the hazard stage, a thunderstorm-wind hazard generated by the hazard module is applied to the power system topology, causing random line failures according to the fragility model. During the restoration stage, repair crews sequentially repair the failed lines based on a predefined repair strategy until full service is restored. Throughout both stages, the system performance—defined as the number of customers served, normalized by the total number of customers—is recorded. The resilience for each event, $Rs_{i,k}$, is computed using the resilience trapezoid method, and the overall system resilience is determined using~(\ref{eq:unweighted_resilience}).

\subsubsection{Power System Topology}

\begin{figure}[thb]
    \centering
    \includegraphics[width=0.8\linewidth]{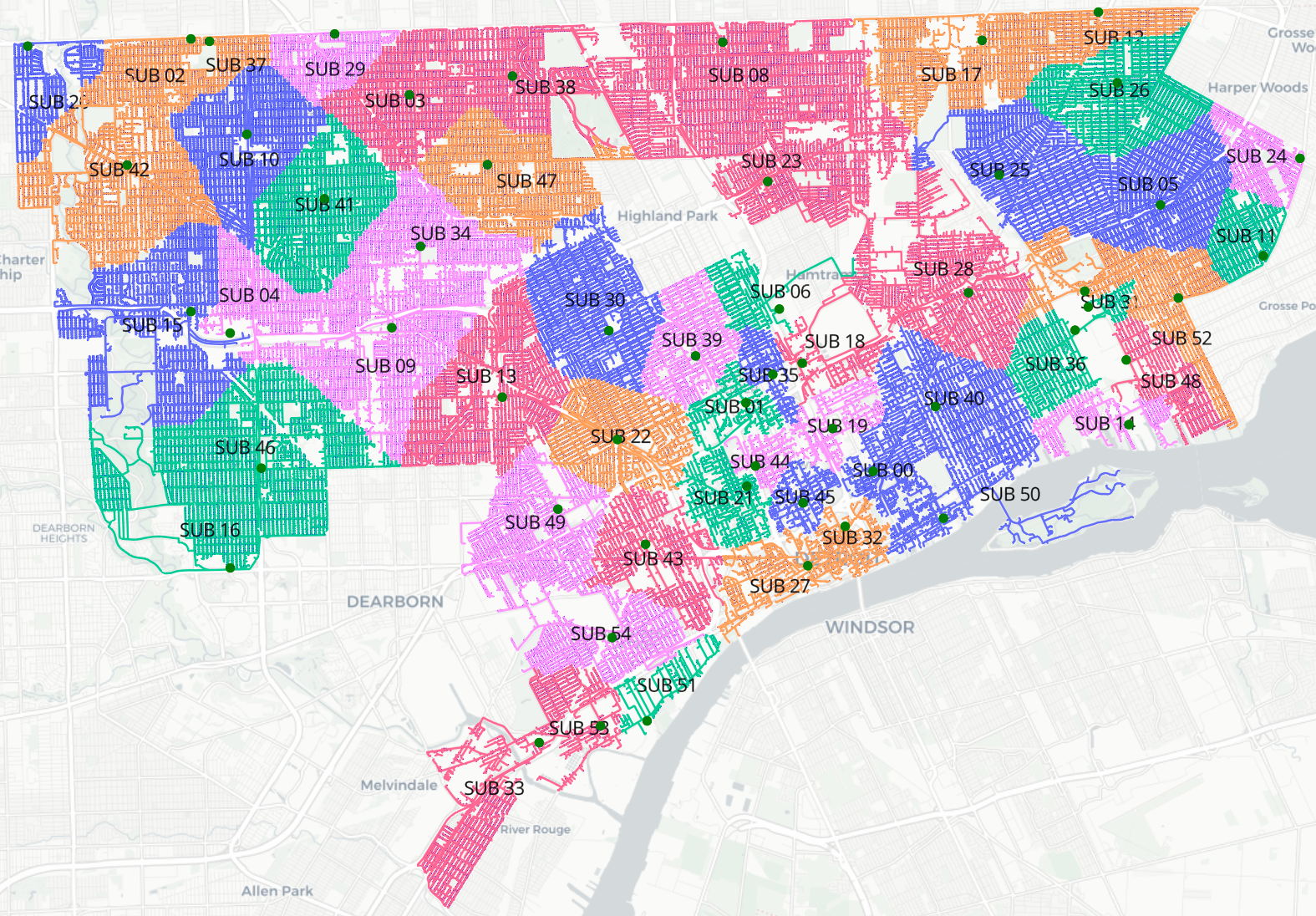}
    \caption{Synthetic power system topology for Detroit generated using publicly available data, composed of poles, lines, substations, and buildings. Green dots are substations. Lines of different colors belong to different substation service areas.}
    \label{fig:power_system_topology}
\end{figure}

Due to security and confidentiality constraints, real distribution network topologies are not publicly available. Therefore, following the approach in~\cite{R71}, a synthetic distribution network for the City of Detroit was generated using publicly accessible datasets, including annual utility reports, OpenStreetMap~\cite{R70}, building footprints, and road maps. The synthetic network, shown in Fig.~\ref{fig:power_system_topology}, consists of 55 substation service areas and four component types: poles, lines, substations, and buildings. The number of customers in each building is assumed to be proportional to its floor area. Each building is connected to a single substation through the shortest available path, and there are no interconnections between substation service areas. A building retains power when a valid electrical connection exists between it and its corresponding substation; when any line along that path is broken, the building—and all its associated customers—loses power.

\subsubsection{Hazard Generator}

The hazard considered in this case study is thunderstorm-induced wind. Only the effect of wind speed on line fragility is modeled; wind direction is neglected. Two wind types are distinguished based on duration and intensity: sustained winds and short-duration gusts. Severe thunderstorms typically involve one or more high-speed gusts that cause localized damage. For each simulated event, the hours and substation service areas affected by wind gusts are predetermined, and line failures are computed based on wind speed at each line's midpoint.

The hazard generator operates in two steps:
\begin{itemize}
    \item A sparse wind speed field is generated for the study area.
    \item The wind speed at each line is obtained by interpolating the sparse field.
\end{itemize}

\begin{figure}[thb]
    \centering
    \includegraphics[width=1.0\linewidth]{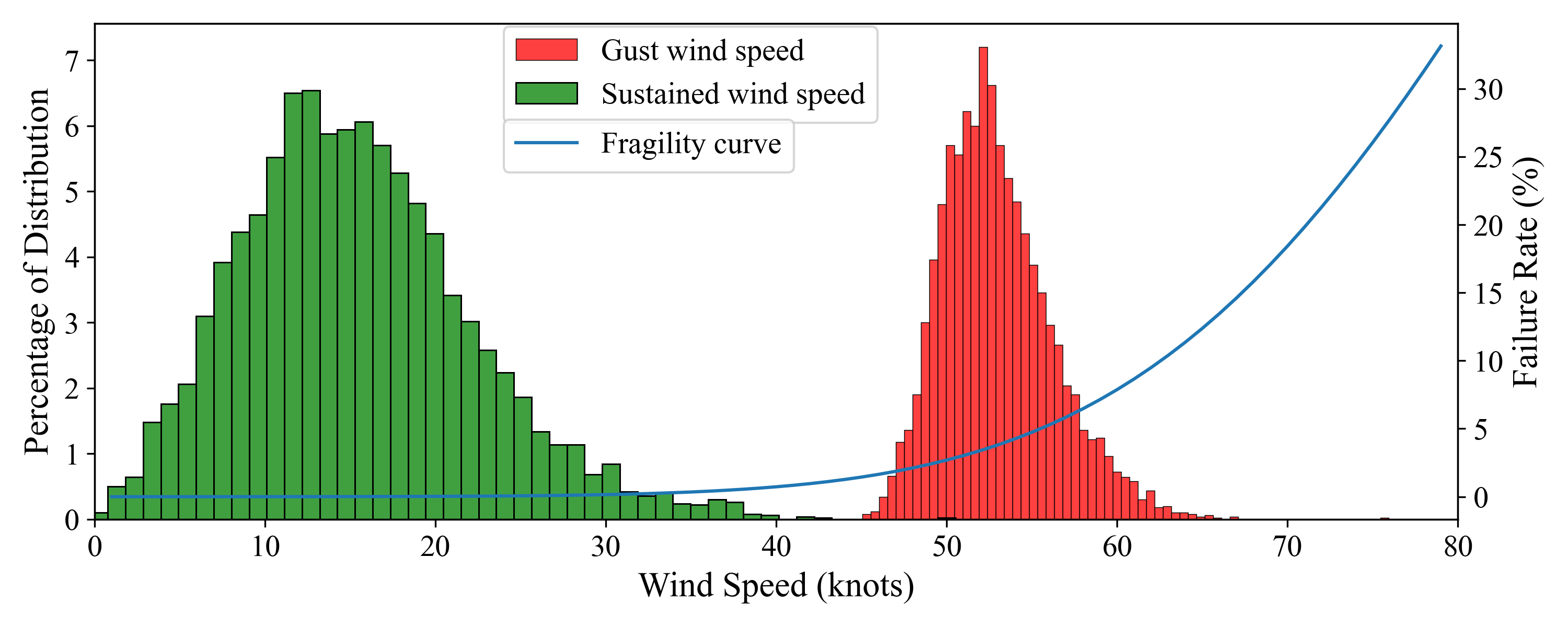}
    \caption{Wind speed distribution and the fragility curve of distribution lines in terms of wind speed.}
    \label{fig:wind_speed_dist_fragility}
\end{figure}

To generate the sparse field, a random location is sampled from each substation service area, and its wind speed is drawn from the corresponding statistical distribution, depending on whether it represents a gust or sustained wind point. The gust distribution is fitted using historical thunderstorm data from Wayne County, Michigan (2010-2023), while the sustained wind distribution is fitted using High-Resolution Rapid Refresh (HRRR)-modeled wind data~\cite{R58,R59,R60} from the same events. The HRRR model, developed by the National Oceanic and Atmospheric Administration (NOAA), provides high-resolution, frequently updated numerical weather forecasts covering the continental United States, Canada, and Mexico. Both fitted distributions are shown in Fig.~\ref{fig:wind_speed_dist_fragility}. To accelerate interpolation, the study area is divided into patches, and the wind speed is assumed constant within each patch.

\subsubsection{Fragility Model}

Two primary failure modes are considered for distribution lines: (1) direct wind-induced damage and (2) damage from fallen tree branches during thunderstorms. Because detailed empirical data on line failure probabilities are unavailable, the fragility curves used in this study are illustrative and designed for demonstration purposes. The wind-speed-based fragility curve for distribution lines is shown in Fig.~\ref{fig:wind_speed_dist_fragility}.

\begin{figure}[thb]
    \centering
    \includegraphics[width=0.8\linewidth]{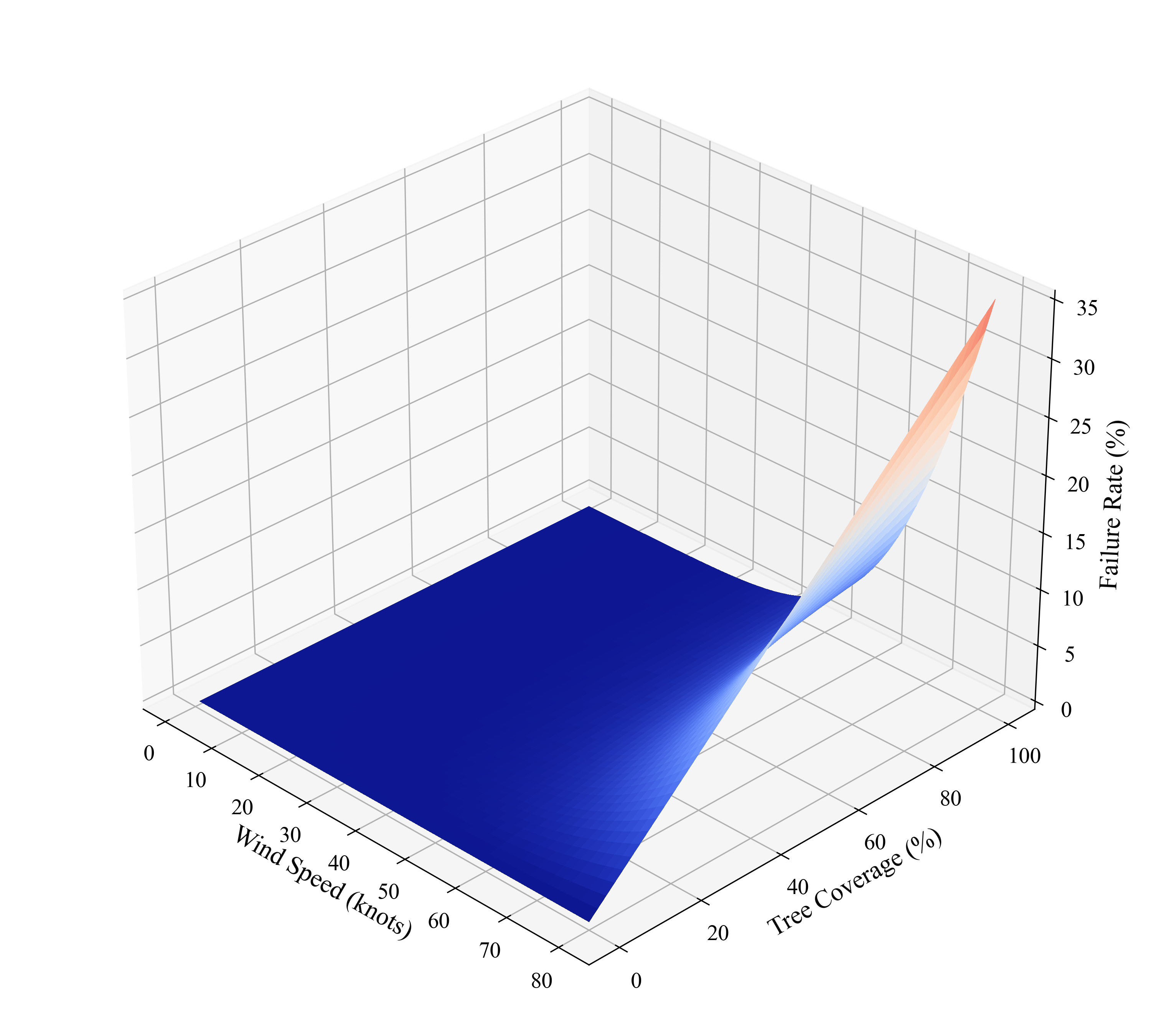}
    \caption{Fragility model of distribution lines of joint impact of wind speed and tree coverage.}
    \label{fig:wind_tree_fragility}
\end{figure}

To account for failures caused by fallen trees, line-specific tree coverage data were obtained from the U.S. Forest Service's public tree canopy dataset\footnote{https://data.fs.usda.gov/geodata/rastergateway/treecanopycover}. The combined wind-tree fragility model is presented in Fig.~\ref{fig:wind_tree_fragility}. Each line is assumed to fail independently according to these fragility relationships. When a line fails, it is removed from the network topology, thereby disconnecting all downstream customers from the supplying substation.

\subsubsection{Recovery Model}

The recovery process models the actions of 15 repair teams shared among all substation service areas. At the beginning of each simulation episode, the initial locations of the repair teams are determined using K-Means clustering of the centroids of the substation service areas. During the hazard stage, no repairs are performed. Once the restoration stage begins, repair teams are dispatched to fix damaged lines according to a priority order based on line criticality—that is, the number of customers who would lose service if the line were removed from the network. Each team spends a random number of hours repairing each failed line, and the restoration process continues until all damaged lines have been repaired. The episode concludes when full service is restored across the system.

\subsubsection{Dataset Organization and Model Training}

The simulated hazard in this case study is thunderstorms, which predominantly occur during the summer season. To obtain a uniform weather representation for each substation service area as the input to the GRU model, the weather input dimension was fixed. All line-center coordinates were clustered using K-Means, and the centroid locations were used to resample the wind speed field. For each substation service area, a feature vector in $\mathbb{R}^{16}$ was extracted as its uniform weather representation. The system embedding was implemented as a one-hot encoding of the substation service area. Note that in this case study the start and end times of each simulated event correspond to the beginning and end of the thunderstorm rather than the outage duration. The resilience of substation service area $k$ under hazard $i$, denoted as $Rs_{i,k}$, was computed using the resilience trapezoid method and served as the ground truth for model training and testing. Only data from substation service areas that experienced wind gusts were used for training and evaluation.

The dataset was split into 80\% for training and validation and 20\% for testing. Stratified splitting was applied so that, for each substation service area, 60\% of the data were used for training, 20\% for validation, and 20\% for testing. On average, each service area contained 422 data samples. All models were trained for 200 epochs, and the model with the lowest validation loss in each fold was selected for testing. Hyperparameters were tuned using the same procedure described in Section~\ref{model_intro_section}. The search space and the optimal values are listed in Table~\ref{table:search_space_B}. The best models achieved a mean absolute error (MAE) of 0.009892 on the validation set and 0.009987 on the test set.

\begin{table}
\begin{center}
\caption{Search Space of the Hyperparameters and the Best Hyperparameters Searched for Case Study A.}
\begin{tabular}{l|l|l}
\hline
Hyperparameter & Search Space & Search Result \\\hline
GRU hidden layer size & [16, 32, 64] & 16 \\
Number of GRU layers & [1, 2, 3, 4] & 4 \\
Number of MLP layers & [1, 2, 3, 4] & 3 \\
GRU dropout rate & 0.0 - 0.8 & 0.0 \\
MLP dropout rate & 0.0 - 0.8 & 0.2 \\
Learning rate & 1e-4 - 1e-2 & 0.00819 \\
Weight decay & 1e-5 - 1e-2 & 0.00002 \\\hline
\end{tabular}
\label{table:search_space_B}
\end{center}
\end{table}

\subsubsection{Resilience Evaluation}

\begin{figure*}[htbp]
\centering
\includegraphics[width=1\linewidth]{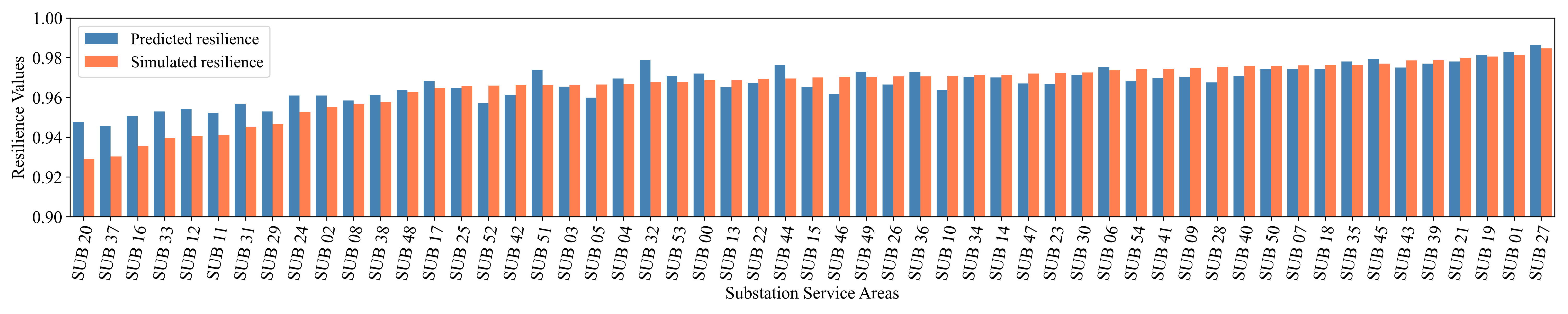}
\vspace*{-3mm}
\caption{Resilience values from simulation results and the model prediction.}
\label{fig:simulated_and_predicted_resilience}
\end{figure*}

The benchmark dataset for this case study comprised all hazardous weather scenarios across the substation service areas in the test set. Each substation service area was evaluated against the benchmark dataset using the trained models, and the average of the test results for each service area was taken as its predicted resilience. The resilience values predicted by the proposed model were then compared with those obtained directly from simulation, as shown in Fig.~\ref{fig:simulated_and_predicted_resilience}.

It is important to note that the weather scenarios used to generate the simulation and prediction results differ. The simulation results were obtained from unique hazard scenarios specific to each substation service area (approximately 422 scenarios each), whereas the resilience predicted by the proposed method was evaluated using the shared benchmark dataset (4,643 weather scenarios in total). Despite this difference, the predicted resilience values closely followed the same trend as the simulated results, with a Spearman rank correlation coefficient of 0.839 ($p = 1.23 \times 10^{-15}$). These results demonstrate that the proposed deep learning model can accurately reproduce the resilience patterns generated by the physical simulation framework, validating its ability to generalize from data without requiring detailed physical models.

\subsection{Case Study B: Application to Real-World Power Outage Data}
\label{section_4_2}

Having validated the proposed framework using simulated data, we next applied it to real-world power outage records to evaluate regional resilience and demonstrate its application in practical planning contexts. Using historical outage and weather data, we estimated each county's resilience (unweighted and weighted), examined how socio-economic and demographic factors shape the weighted metric, and demonstrated a simple DER sizing exercise to meet specified resilience targets.

\subsubsection{Data Collection and Preprocessing}

To train the deep learning model, power outage data from the EAGLE-I dataset~\cite{R52} were used to generate the ground-truth resilience values $Rs_{i,k}$. All reported power outages in Michigan were extracted and downsampled from 15-minute intervals to hourly resolution. Three counties were excluded because their first recorded outages occurred in December 2019, suggesting they were added to the EAGLE-I system later than others. Following the approach in~\cite{R37}, assuming an average household size of two persons, the outage counts were converted to estimated numbers of affected individuals using a scale factor of 2, and the resulting values were normalized by the county population in the corresponding year.

The goal of this study is to estimate system resilience under low-probability, high-impact (LPHI) events rather than to perform short-term outage prediction. Accordingly, all normalized outage values below 0.1 were filtered out, retaining only hours when at least 10\% of the population lost power. When the gap between two consecutive outage periods within the same county was less than three hours, the two periods were merged, and missing values were linearly interpolated. Outage events lasting less than six hours were removed. Counties with fewer than two valid events were excluded to ensure representation in both the training and validation sets. This process resulted in 71 Michigan counties with a total of 682 outage events between 2014 and 2022. The corresponding $Rs_{i,k}$ values were calculated using the resilience trapezoid and served as the ground truth for model training.

\paragraph{Weather Data}
The weather data were obtained from the HRRR model outputs~\cite{R58,R59,R60} using the Herbie interface~\cite{R61}. For each outage event, 18 surface variables were extracted at the population center of each county at hourly resolution. These variables include wind gust, precipitation rate, snow depth, snow cover, pressure, temperature, total precipitation, visibility, and others. Missing values were filled using the nearest available timestamp, and all variables were normalized to the range $[0, 1]$.

\paragraph{Socio-Economic and Demographic Data}
Most socio-economic and demographic factors used for computing the weighted resilience were derived from the U.S. Census Bureau's American Community Survey (ACS) 5-Year Estimates for 2022. Data on nursing home residents were obtained from the 2020 Decennial Census, and information on residents dependent on electricity-powered medical equipment was sourced from the U.S. Department of Health and Human Services (HHS) emPOWER program~\cite{R63}. The 15 factors and their corresponding data sources are summarized in Table~\ref{table:weight_factors}. Each factor was normalized by either the total number of households or the total population of the corresponding county, depending on its definition.

\subsubsection{Model Training and Validation}

A five-fold cross-validation was conducted during model training, and the models with the best validation performance were saved and used for subsequent resilience evaluations. For simplicity, each county was represented by a one-hot encoded system embedding. The (GRU)~\cite{R62} was selected as the recurrent architecture. All models were trained for 500 epochs using the MAE loss function and optimized with Adam~\cite{R64}.

\begin{figure}[tbh]
    \centering
    \includegraphics[width=0.8\linewidth]{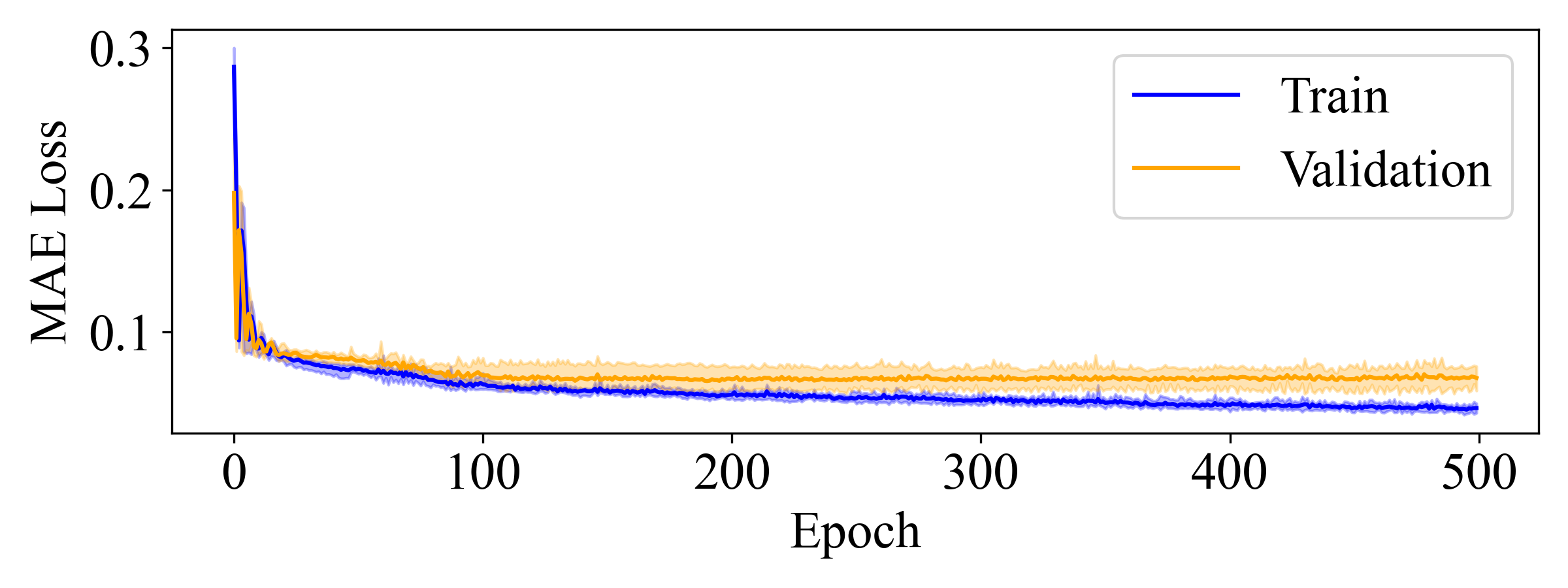}
    \caption{Training and validation losses of case study B, averaged over the 5 folds.}
    \label{fig:loss}
\end{figure}

Hyperparameters were tuned using Ray Tune~\cite{R66} and Optuna~\cite{R67} to identify the best configuration. The search space and the optimal parameters found are listed in Table~\ref{table:search_space_A}. All subsequent analyses were conducted using the best-performing hyperparameter settings. The average training and validation losses across the five folds are shown in Fig.~\ref{fig:loss}. The mean validation loss of the best models across all folds was 0.06301.

\begin{table}[tbh]
\begin{center}
\caption{Search Space of the Hyperparameters and the Best Hyperparameters Searched for Case Study B.}
\begin{tabular}{l|l|l}
\hline
Hyperparameter & Search Space & Search Result \\\hline
GRU hidden layer size & [32, 64, 128, 256] & 128 \\
Number of GRU layers & [1, 2, 3, 4] & 1 \\
Number of MLP layers & [1, 2, 3, 4] & 2 \\
GRU dropout rate & 0.0 - 0.8 & 0.0 \\
MLP dropout rate & 0.0 - 0.8 & 0.2 \\
Learning rate & 1e-4 - 1e-2 & 0.00623 \\
Weight decay & 1e-5 - 1e-2 & 0.00058 \\\hline
\end{tabular}
\label{table:search_space_A}
\end{center}
\end{table}

\subsubsection{Resilience Evaluation}

\begin{figure}[htbp]
    \centering
    \includegraphics[width=0.9\linewidth]{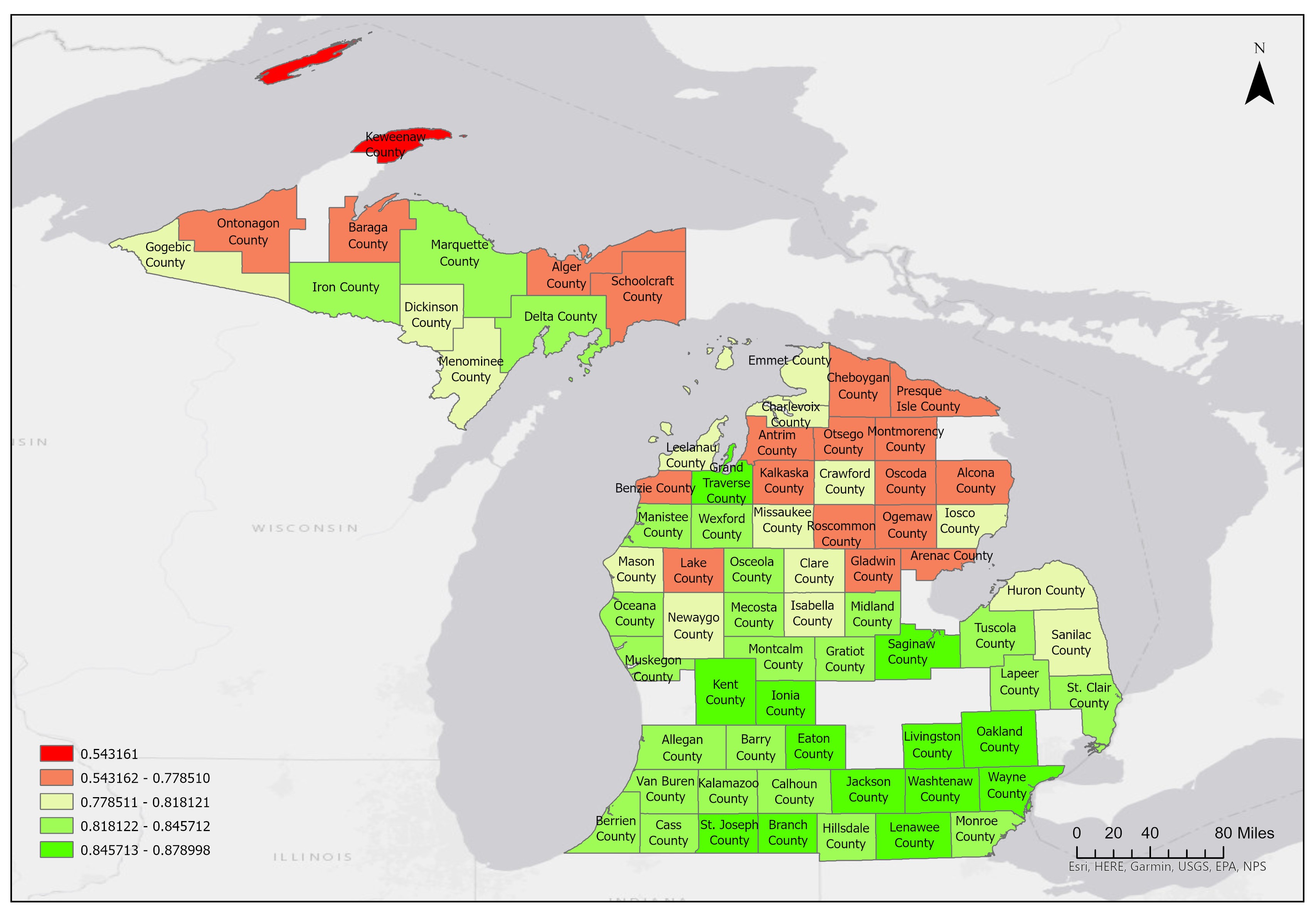}
    \caption{Unweighted electric power resilience of 71 counties in Michigan, USA.}
    \label{fig:unweighted_resilience_results}
\end{figure}

\begin{figure}[htbp]
    \centering
    \includegraphics[width=0.9\linewidth]{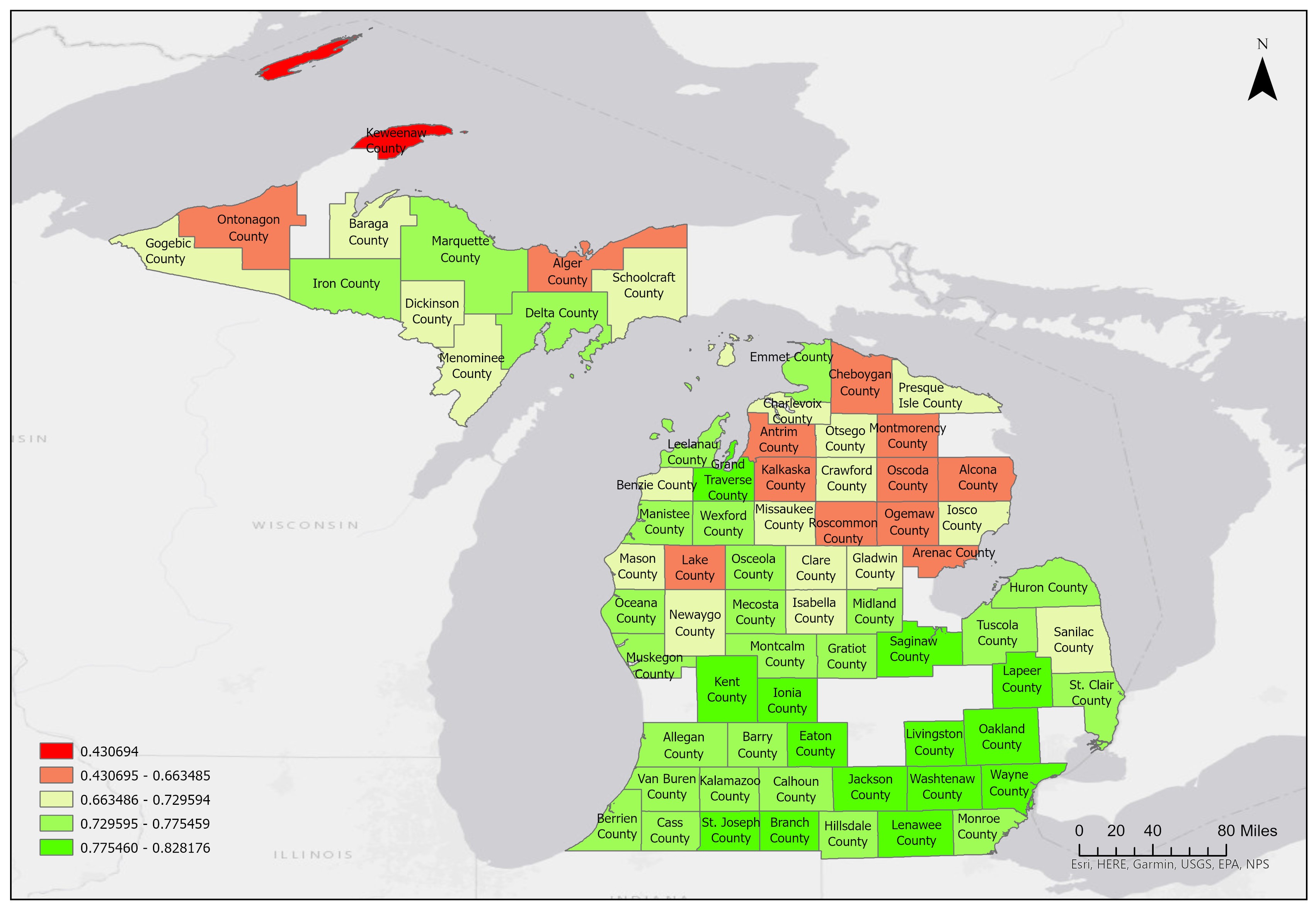}
    \caption{Weighted electric power resilience of 71 counties in Michigan, USA.}
    \label{fig:weighted_resilience_results_1}
\end{figure}

\begin{figure*}[htbp]
\centering
\includegraphics[width=1\textwidth]{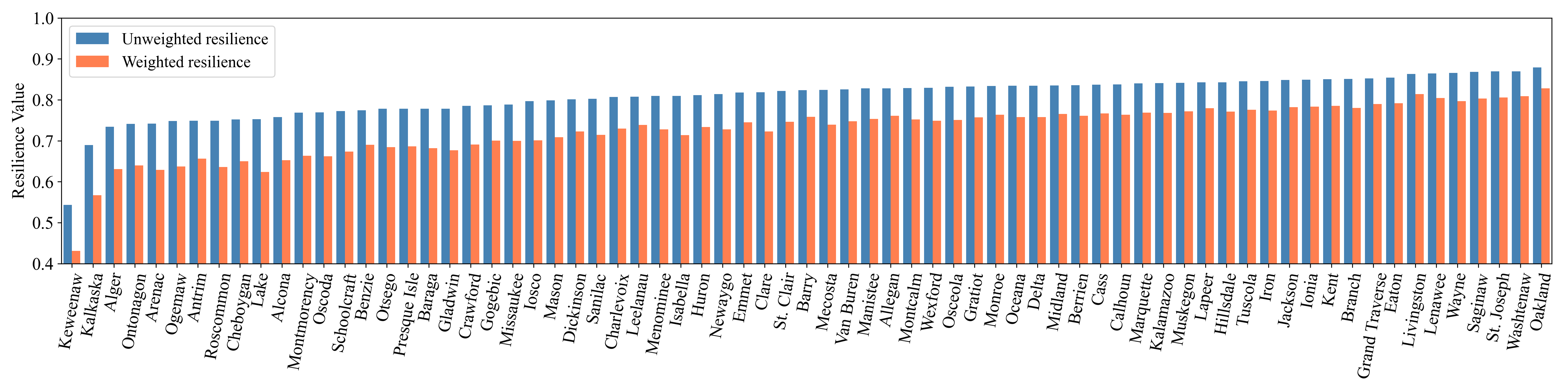}
\vspace*{-3mm}
\caption{Unweighted and weighted electric power resilience of 71 counties in Michigan, USA.}
\label{fig:unweighted_and_weighted_results}
\end{figure*}

A benchmark dataset of hazardous weather scenarios was constructed using all weather events from 2020 to 2022 across the 71 Michigan counties, totaling 370 events. Each county was evaluated against this benchmark dataset using the trained model, and both unweighted and weighted resilience were computed using~(\ref{eq:unweighted_resilience}) and~(\ref{eq:weighted_resilience}), respectively. The resulting unweighted and weighted resilience maps are shown in Fig.~\ref{fig:unweighted_resilience_results} and Fig.~\ref{fig:weighted_resilience_results_1}. The corresponding county-level rankings are compared in Fig.~\ref{fig:unweighted_and_weighted_results}. The Spearman rank-order correlation coefficient between the unweighted and weighted resilience values is 0.99 ($p = 8 \times 10^{-62}$), indicating that incorporating the weighting terms did not substantially alter the overall ranking of county resilience.

\begin{figure*}[htbp]
\centering
\includegraphics[width=1\textwidth]{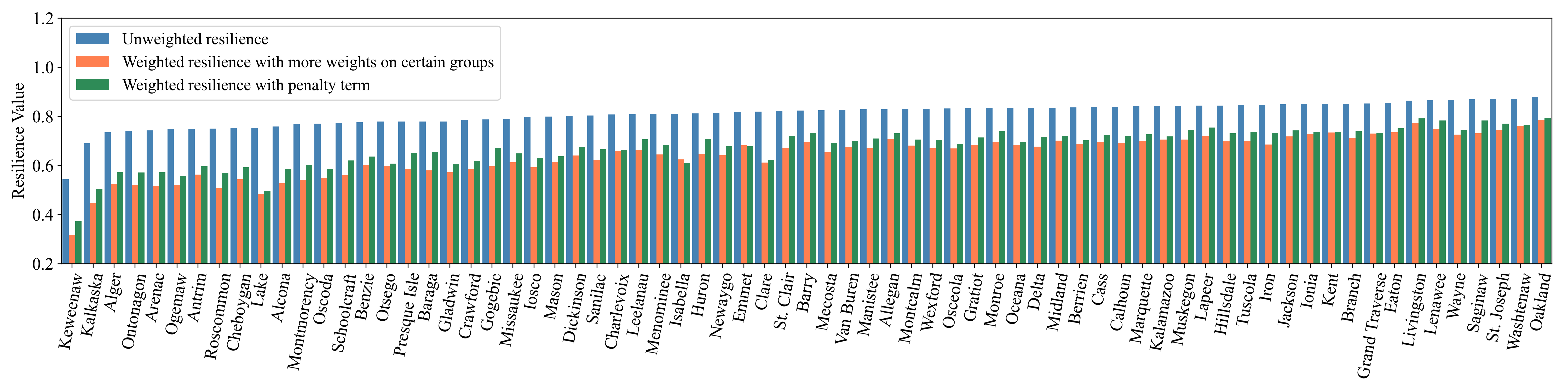}
\vspace*{-3mm}
\caption{Two examples of weighting methods. 1) weighted resilience with more weights on three population groups. 2) weighted resilience with more weights on demographic indicators larger than 0.2.}
\label{fig:weighted_more_examples}
\end{figure*}

As discussed in Section~\ref{section_3_1}, the weighting terms can be adjusted to reflect different policy priorities or to emphasize specific demographic groups. To illustrate this flexibility, two alternative weighting schemes were evaluated. In the first, greater weights (5x) were assigned to three population groups—people with disabilities, elderly individuals living alone, and low-income households. In the second, following the approach in~\cite{R5}, a penalty factor of 3 was applied to any demographic indicator exceeding 0.2, reflecting that regions with highly concentrated vulnerable populations should be considered less resilient. The results for these two examples are shown in Fig.~\ref{fig:weighted_more_examples}. The Spearman correlation coefficients between the unweighted and weighted resilience under these two schemes are 0.97 ($p = 3 \times 10^{-45}$) and 0.95 ($p = 1 \times 10^{-37}$), respectively. These examples demonstrate that different weighting strategies can meaningfully adjust the relative resilience rankings. For instance, St.~Joseph County ranks lower than Washtenaw County in the first example but higher in the second. Such changes highlight how the proposed framework enables policymakers to explore the impact of alternative weighting approaches when prioritizing investments for resilience enhancement.

\begin{figure*}[htbp]
\centering
\includegraphics[width=1\textwidth]{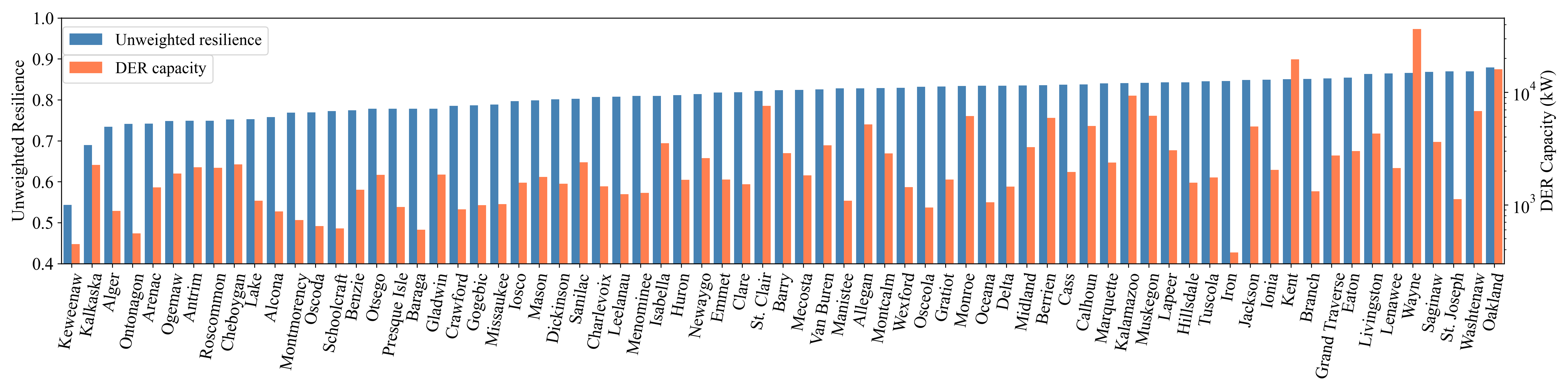}
\vspace*{-3mm}
\caption{Minimum capacity of DERs to be installed to enhance the unweighted resilience to 0.9 for each county.}
\label{fig:planning_unweighted_results}
\end{figure*}

\begin{figure*}[htbp]
\centering
\includegraphics[width=1\linewidth]{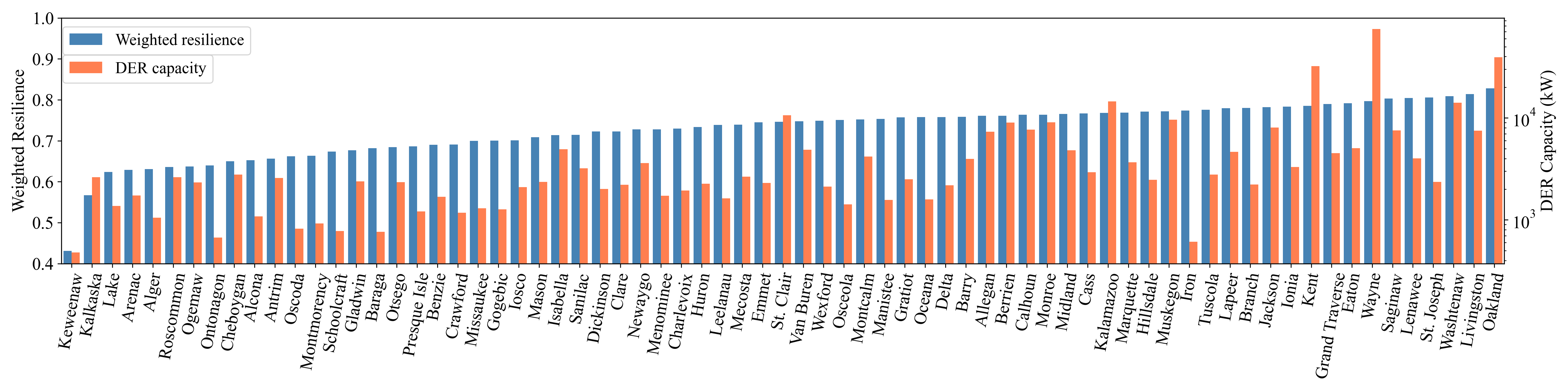}
\vspace*{-3mm}
\caption{Minimum capacity of DERs to be installed to enhance the weighted resilience to 0.9 for each county.}
\label{fig:planning_weighted_results}
\end{figure*}

\subsubsection{Resilience Enhancement}

The resilience evaluation results can further support resilience enhancement planning by estimating the distributed energy resource (DER) capacity required to achieve specific resilience targets. Assuming that DERs can provide backup power during hazardous weather events, the minimum capacity needed can be determined by solving~(\ref{eq:planning_unweighted}) or~(\ref{eq:planning_weighted}), given a target unweighted or weighted resilience, respectively. In these equations, $\overline{Ru}_k$ and $\overline{Rw}_k$ represent the target unweighted and weighted resilience values of system $k$, $P_k$ denotes the minimum DER capacity to be installed, $Pu$ is the average power demand per customer, and $Np_k$ is the number of customers in region $k$. Following~\cite{R68}, the average demand per customer was set to $Pu = 600$~W.

As an illustrative example, the minimum DER capacity required for each county to reach a resilience target of 0.9 is shown in Fig.~\ref{fig:planning_unweighted_results} (unweighted) and Fig.~\ref{fig:planning_weighted_results} (weighted). The Spearman rank correlation coefficients between the computed resilience values and the required DER capacities are 0.53 ($p = 1.55 \times 10^{-6}$) and 0.66 ($p = 4.2 \times 10^{-10}$), respectively. These correlations indicate that while more resilient counties generally require less DER capacity to achieve the same target, population size also strongly influences the required capacity. For example, Wayne County, which ranks 5th and 7th in unweighted and weighted resilience, respectively, demands the largest DER capacity due to its large population. In real-world applications, additional factors such as regional demand profiles and DER dispatch strategies could further alter these rankings.

\begin{equation}
    \label{eq:planning_unweighted}
    \overline{Ru}_k=\frac{1}{N_a}\sum_{i=1}^{N_a}(Rs_{i,k} + \frac{P_k}{PuNp_k})
\end{equation}

\begin{equation}
    \label{eq:planning_weighted}
    \overline{Rw}_k=[\frac{1}{N_a}\sum_{i=1}^{N_a}(Rs_{i,k} + \frac{P_k}{PuNp_k})]^{[1+\frac{1}{N_w}\sum_{i=1}^{N_w}W_{j,k}]}
\end{equation}

\section{Discussion}
\label{section_5}

While the proposed deep learning-based framework provides a flexible and scalable approach to evaluating power system resilience, several limitations and practical considerations merit discussion:

\begin{itemize}
    \item \textbf{Data dependency.} Model performance and generalizability depend on the availability, coverage, and fidelity of historical outage and weather data. In regions where records are sparse, delayed, or noisy, predictive accuracy may degrade and uncertainty may increase, especially for rare extreme events.

    \item \textbf{Use without detailed topology.} A key strength of the framework is its applicability when detailed network topology and component fragility models are unavailable, making it useful for planners and policymakers with limited engineering datasets. When high-quality physical models are accessible, however, simulation-based engineering studies can yield more granular, operationally actionable insights (e.g., component-level interventions).

    \item \textbf{Implicit recovery modeling.} Recovery dynamics are learned implicitly from historical performance curves. This assumes broadly consistent restoration practices and resource levels over time. Significant changes in crew allocation, mutual-assistance policies, or restoration priorities may induce shift and bias the learned mapping.

    \item \textbf{Social-vulnerability weighting.} The weighting mechanism based on normalized demographic and socioeconomic indicators represents one defensible formulation. Alternative formulations may be preferable where local priorities call for emphasizing specific population groups or enforcing policy constraints; the framework is compatible with such adjustments.

    \item \textbf{Planning resolution for DERs.} The planning example estimates aggregate DER capacity to meet resilience targets. It does not optimize siting, sizing, or dispatch at the feeder or node level, which would require explicit network models and spatiotemporal load detail.
\end{itemize}

These limitations suggest several avenues for future work. Integrating the framework with physical network models (to support location-specific planning), enhancing interpretability of learned weather-impact relationships, quantifying predictive uncertainty under distribution shift, and extending the approach to dynamic DER siting and operation are promising directions that could improve both strategic and operational decision-making.

\section{Conclusion}
\label{section_6}

This work introduced a deep learning-based method for evaluating power system resilience that bridges statistics-based and simulation-based approaches. The model learns the relationship between hazardous weather inputs and system performance and directly predicts event-level resilience, which is then aggregated over a common benchmark of weather scenarios to yield comparable system-level metrics. When desired, socio-economic and demographic factors can be incorporated to produce a weighted resilience measure that reflects policy priorities regarding vulnerable populations.

Two case studies demonstrated the method's effectiveness and use cases. First, a controlled validation with synthetic data generated by a graph-based simulation framework showed that the model reproduces simulated resilience patterns with strong rank agreement. Second, an application to real outage and weather data at the county level in Michigan, USA produced statewide resilience estimates and illustrated how the results can inform planning, including estimating DER capacity needed to meet target resilience levels. Together, these results indicate that the proposed framework offers a scalable, data-driven basis for comparative resilience assessment and policy-relevant planning without requiring detailed physical models.





\bibliographystyle{elsarticle-num}
\bibliography{ref}






\end{document}